\documentclass[13pt,a4paper,indentfirst]{extarticle}

\usepackage{geometry}
\usepackage{rotating}
\usepackage{longtable}
\usepackage{amsmath}
\usepackage{graphicx} 
\usepackage{dcolumn}  
\usepackage{epsfig}

\usepackage{booktabs}
\usepackage{multirow}

\renewcommand{\arraystretch}{1.1}

\newcommand{\kev}{\mathrm{keV}}
\newcommand{\mev}{\mathrm{MeV}}

\newcommand{\mevm}{\mathrm{MeV}/c^2}
\newcommand{\gev}{\mathrm{GeV}}

\newcommand{\gevm}{\mathrm{GeV}/c^2}

\newcommand{\ee}{e^+e^-}
\newcommand{\uu}{\mu^+\mu^-}
\newcommand{\pp}{\pi^+\pi^-}
\newcommand{\kk}{K^+K^-}

\newcommand{\U}{\Upsilon}
\newcommand{\Ufo}{\Upsilon(4S)}
\newcommand{\Uf}{\Upsilon(5S)}
\newcommand{\Us}{\Upsilon(6S)}

\newcommand{\Uo}{\Upsilon(1S)}

\newcommand{\Ut}{\Upsilon(2S)}
\newcommand{\Uth}{\Upsilon(3S)}
\newcommand{\Un}{\Upsilon(nS)}

\newcommand{\et}{\eta_b(1S)}
\newcommand{\ett}{\eta_b(2S)}

\newcommand{\hb}{h_b(1P)}
\newcommand{\hbp}{h_b(2P)}
\newcommand{\hbn}{h_b(nP)}
\newcommand{\hbm}{h_b(mP)}
\newcommand{\jp}{J/\psi}

\newcommand{\psp}{\psi(2S)}

\newcommand{\pb}{\mathrm{pb}^{-1}}
\newcommand{\fb}{\mathrm{fb}^{-1}}

\newcommand{\zbo}{Z_b(10610)}
\newcommand{\zbt}{Z_b(10650)}

\begin{document}
\begin{flushright}
{\small FTPI-MINN-16/28\\ 
UMN-TH-3606/16}
\end{flushright}

\begin{center}
{\Large \bf\boldmath Bottomonium-like states: physics case for
  energy scan above the $B\bar{B}$ threshold at Belle-II}
\vskip 5mm
A.~E.~Bondar$^{a,b}$, R.~V.~Mizuk$^{c,d}$ and M.~B.~Voloshin$^{e,f,g}$
\vskip 3mm
{\small {\it $^a$ Budker Institute of Nuclear Physics, Novosibirsk, 630090, Russia}}\\
{\small {\it $^b$ Novosibirsk State University, Novosibirsk, 630090, Russia}}\\
{\small {\it $^c$ Lebedev Physical Institute of the Russian Academy of Sciences, Moscow, 119991, Russia}}\\
{\small {\it $^d$ National Research Nuclear University MEPhI (Moscow Engineering Physics Institute), Moscow, 115409, Russia}}\\
{\small {\it $^e$ William I. Fine Theoretical Physics Institute, University of
Minnesota, Minneapolis, MN 55455, USA}}\\
{\small {\it $^f$ School of Physics and Astronomy, University of Minnesota, Minneapolis, MN 55455, USA}}\\
{\small {\it $^g$ Institute of Theoretical and Experimental Physics, Moscow, 117218, Russia}}\\
\end{center}

\begin{abstract}
The Belle-II experiment is expected to collect large data samples at
the $\Ufo$ and $\Uf$ resonances to study primarily $B$ and $B_s$
mesons. We discuss what other data above the $B\bar{B}$ threshold are
of interest. We propose to perform a high-statistics energy scan from
the $B\bar{B}$ threshold up to the highest possible energy, and to
collect data at the $\Us$ and at higher mass states if they are found in
the scan.
We emphasize the interest in increasing the maximal energy from
$11.24\,\gev$ to $11.5-12\,\gev$ in the future.
These data are needed for investigation of bottomonium and
bottomonium-like states.
\end{abstract}

\section{Introduction}

Conventional bottomonium is an approximately non-relativistic
system. Out of 34 expected $b \bar b$ energy levels below the $B\bar{B}$
threshold~\cite{Godfrey:2015dia,Wurtz:2015mqa} 15 have been
observed~\cite{PDG14}. The masses and decays of these states are well
described by potential models~\cite{Brambilla:2004wf}.

There are five states with $b\bar{b}$ pairs above the $B\bar{B}$
threshold: three isospin-zero vector states $\U(10580)$, $\U(10860)$ and
$\U(11020)$ [or $\Ufo$, $\Uf$ and $\Us$ according to the potential
  model assignment], and two isospin-one axial states $\zbo$ and
$\zbt$. The isospin-one states are obviously exotic with minimal quark
content $|b\bar{b}u\bar{d}\rangle$. But the isospin-zero states
also have properties unexpected for a pure $b\bar{b}$ pair. The mass
splitting between the $\Ufo$ and $\Uf$ is larger by $73\pm11\,\mevm$ than
that between the $\Uth$ and $\Ufo$, while for a pure $b\bar{b}$ system it is
expected to be smaller by about
$40\,\mevm$~\cite{Godfrey:2015dia}. The rates of $\Uf\to\Un\pp$ and
$\Us\to\Un\pp$ transitions are two orders of magnitude higher than
expected for a pure bottomonium~\cite{Abe:2007tk,Santel:2015qga}. The
$\eta$ transitions, that in a pure bottomonium involve spin of heavy
quark and are suppressed by three orders of magnitude relative to
$\pp$ transitions, are not strongly suppressed in case of
the $\Uf$~\cite{Krokovny:LaThuile2012} and are even enhanced in case of
the $\Ufo$~\cite{Aubert:2008az}. More puzzling results on
hadronic transitions are discussed below. Thus, there is a strong
evidence that the structure of the vector states above the $B\bar{B}$
threshold is more complicated than the pure $b\bar{b}$, i.e.\ it is
exotic.

In this paper we briefly review the experimental status of the
``botomonium-like'' states, their interpretation and prospects for
future studies. Our main goal is to discuss what we can learn about
them based on the increased $\Ufo$ and $\Uf$ data samples at Belle-II,
and what dedicated data taking is needed. We propose to perform a
high-statistics energy scan from the $B\bar{B}$ threshold up to the
highest possible energy to measure exclusive open and hidden flavor
cross sections. We also propose to collect data at the peak of $\Us$
and at higher mass states -- if they are found in the energy scan --
to study their properties and to search for missing bottomonia,
$P$-wave $B_s$ mesons or exotic states -- partners of the $\zbo$ and
$\zbt$. 

Favored interpretation of the bottomonium-like states is related to
the presence of heavy mesons in their wave functions. The $\zbo$ and
$\zbt$ states have purely molecular structures, $B\bar{B}^*$ and
$B^*\bar{B}^*$, respectively, while the $\Ufo$, $\Uf$ and $\Us$ states
are mixtures of the $b\bar{b}$ and
$B_{(s)}^{((*)*)}\bar{B}_{(s)}^{((*)*)}$ pairs [with $B_{(s)}^{**}$ we
denote $P$-wave excitations of $B$ or $B_s$ mesons]. We discuss
measurements that are needed to establish this interpretation and
point out strategy to search for other types of exotic hadrons --
compact tetraquarks~\cite{Ali:2009es} or
hadroquarkonia~\cite{Dubynskiy:2008mq}.

The paper is organized as follows. We first discuss the results
obtained using on-resonance data, since they constitute the bulk of
our current knowledge about the bottomonium-like states, and then move
to the discussion of the energy scans. We conclude with a summary of
the physics program for the proposed dedicated data taking.

\section{On-resonance data}

The results on bottomonium-like states presented in this section are
based on large data samples collected by the Belle and BaBar
experiments. These are $711\,\fb$ and $433\,\fb$ collected by Belle
and BaBar, respectively, at the $\Ufo$, and $121\,\fb$ collected by Belle
at the $\Uf$. In the $\Us$ region only scan data are available -- five points
with a total luminosity of $5\,\fb$ that correspond to $3\,\fb$ collected
in the peak.

\subsection{Mechanism of hadronic transitions and structure of vector
  bottomonium-like states}

Known hadronic transitions from vector bottomonium and
bottomonium-like states and corresponding partial widths are presented
in Table~\ref{tab:partial_widths}\footnote{We do not list transitions
  to the $\hbn\pp$ final states, since they proceed entirely via the
  intermediate $Z_b$
  states~\cite{Adachi:2011ji,Abdesselam:2015zza}. We do not subtract
  the $Z_b$ contributions from the $\Un\pp$ final states; they should
  be relatively small in all transitions except $\Us\to\Uth\pp$.}.
\begin{table}[tbh]
\caption{ Known hadronic transitions from vector bottomonium(-like)
  states and corresponding partial widths. } \centering
\begin{tabular}{lcc}
\hline\hline
Transition & Partial width (keV) & Reference \\
\hline
$\Ut\to$  &  &  \\
\hspace{5mm}$\Uo\,\pp$  & $5.7\pm0.5$ & \cite{PDG14} \\
\hspace{5mm}$\Uo\,\eta$ & $(9.3\pm1.5)\times10^{-3}$ & \cite{PDG14} \\
\hline
$\Uth\to$  &  &  \\
\hspace{5mm}$\Uo\,\pp$  & $0.89\pm0.08$ & \cite{PDG14} \\
\hspace{5mm}$\Uo\,\eta$ & $<2\times10^{-3}$ & \cite{PDG14} \\
\hspace{5mm}$\Ut\,\pp$  & $0.57\pm0.06$ & \cite{PDG14} \\
\hline
$\Ufo\to$  &  &  \\
\hspace{5mm}$\Uo\,\pp$  & $1.7\pm0.2$ & \cite{Aubert:2008az,Sokolov:2009zy} \\
\hspace{5mm}$\Uo\,\eta$ & $4.0\pm0.8$ & \cite{Aubert:2008az} \\
\hspace{5mm}$\Ut\,\pp$  & $1.8\pm0.3$ & \cite{Aubert:2008az} \\
\hspace{5mm}$\hb\,\eta$ & $45\pm7$ & \cite{Tamponi:2015xzb} \\
\hline
$\Uf\to$  &  &  \\
\hspace{5mm}$\Uo\,\pp$  & $238\pm41$ & \cite{Garmash:2014dhx} \\
\hspace{5mm}$\Uo\,\eta$ & $39\pm11$ & \cite{Krokovny:LaThuile2012} \\
\hspace{5mm}$\Uo\,\kk$  & $33\pm11$ & \cite{Abe:2007tk} \\
\hspace{5mm}$\Ut\,\pp$  & $428\pm83$ & \cite{Garmash:2014dhx} \\
\hspace{5mm}$\Ut\,\eta$ & $204\pm44$ & \cite{Krokovny:LaThuile2012} \\
\hspace{5mm}$\Uth\,\pp$ & $153\pm31$ & \cite{Garmash:2014dhx} \\
\hspace{5mm}$\chi_{b1}(1P)\,\omega$ & $84\pm20$ & \cite{He:2014sqj} \\
\hspace{5mm}$\chi_{b1}(1P)\,(\pp\pi^0)_\mathrm{non\textnormal{-}\omega}$ & $28\pm11$ & \cite{He:2014sqj} \\
\hspace{5mm}$\chi_{b2}(1P)\,\omega$ & $32\pm15$ & \cite{He:2014sqj} \\
\hspace{5mm}$\chi_{b2}(1P)\,(\pp\pi^0)_\mathrm{non\textnormal{-}\omega}$ & $33\pm20$ & \cite{He:2014sqj} \\
\hspace{5mm}$\U_J(1D)\,\pp$  & $\sim60$ & \cite{Mizuk:QCHS2012} \\
\hspace{5mm}$\U_J(1D)\,\eta$ & $150\pm48$ & \cite{Tamponi:QWG2014} \\
\hspace{5mm}$\zbo^\pm\pi^\mp$ & $2070\pm440$ & \cite{Garmash:2015rfd} \\
\hspace{5mm}$\zbt^\pm\pi^\mp$ & $1200\pm300$ & \cite{Garmash:2015rfd} \\
\hline
$\Us\to$  &  &  \\
\hspace{5mm}$\Uo\,\pp$  & $137\pm32$ & \cite{Santel:2015qga} \\
\hspace{5mm}$\Ut\,\pp$  & $183\pm43$ & \cite{Santel:2015qga} \\
\hspace{5mm}$\Uth\,\pp$ & $77\pm28$ & \cite{Santel:2015qga} \\
\hspace{5mm}$Z_b(10610,10650)^\pm\pi^\mp$ & $1300-6600$ & \cite{Abdesselam:2015zza} \\
\hline\hline
\end{tabular}
\label{tab:partial_widths}
\end{table}
There are difficulties with the determination of the $\Uf$ and $\Us$
partial widths, since, in particular, the corresponding $\ee$ widths
are not yet known. Thus the partial widths are defined as
$\sigma^\mathrm{vis}/\sigma_{b\bar{b}}\times\Gamma$, where
$\sigma^\mathrm{vis}$ is the visible cross section of the considered
process, $\sigma_{b\bar{b}}$ is the total $b\bar{b}$ cross section and
$\Gamma$ is the width of the $\Uf$ or $\Us$. This definition likely
underestimates the partial widths~\cite{def_partial_width}.

\begin{figure}[htbp]
\includegraphics[width=0.46\linewidth]{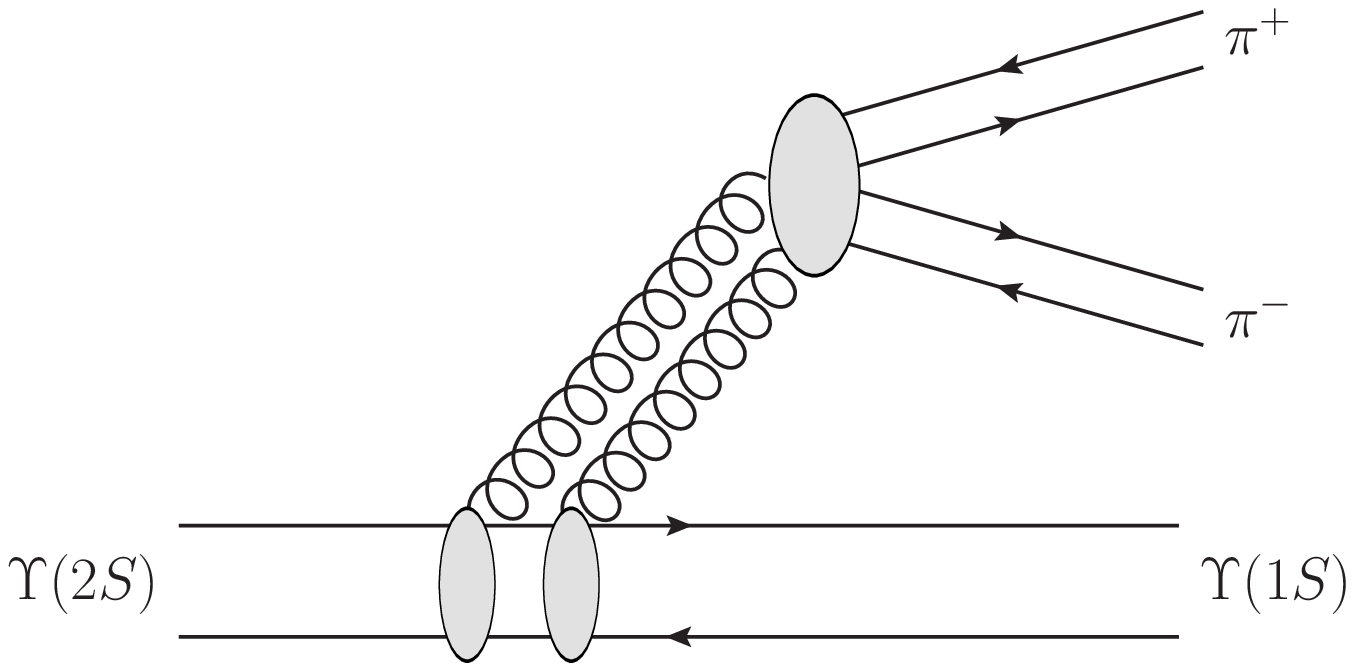}\hfill
\includegraphics[width=0.41\linewidth]{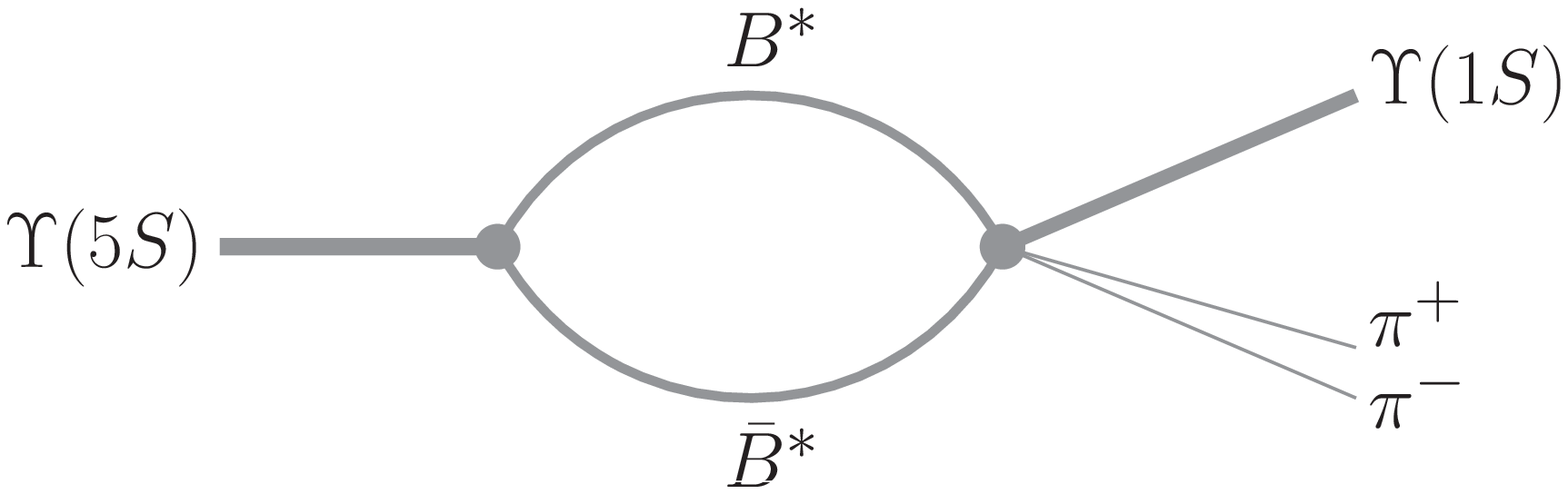}
\caption{ Diagrams of the transitions $\Ut\to\Uo\pp$ (left) and
  $\Uf\to\Uo\pp$ (right). }
\label{fig:diagrams}
\end{figure}

In a pure $b \bar b$ bottomonium the hadronic transitions proceed via
emission of gluons, which convert into light hadrons (see
Fig.~\ref{fig:diagrams}~(left)).
Such transitions are OZI suppressed, accordingly, their partial widths
are small.
If the size of the quarkonium is small compared to the gluon wave
lengths, the transitions can be described using the QCD multipole
expansion~\cite{Gottfried:1977gp,Voloshin:1978hc}. The $\pi^+\pi^-$
transitions between the $\Un$ occur via emission of two $E1E1$ gluons,
while $\eta$ transitions require emission of $E1M2$ gluons. Such
magnetic transitions involve the spin of the heavy $b$ quark and are
strongly suppressed, which is a manifestation of the Heavy Quark Spin
Symmetry (HQSS). These expectations are well fulfilled for the $\Ut$
and $\Uth$ states, which are below the $B\bar{B}$ threshold.

For the states above the $B\bar{B}$ threshold the pattern of
transitions changes. In case of $\Ufo$ the $\eta$ transition to $\Uo$
is not suppressed, but rather enhanced by a factor of $2.4\pm0.4$
relative to the $\pp$ transition, which corresponds to a very strong
violation of HQSS.  Similarly, the $\Ufo\to\hb\eta$ transition in the
multipole expansion corresponds to emission of HQSS-breaking $E1M1$
gluons, but it is found to be even further enhanced. In case of the $\Uf$
and $\Us$, the partial widths are at the level of hundred $\kev$,
which is two orders of magnitude higher than expected for a pure
bottomonium, and is a clear violation of the OZI rule. The $\eta$
transitions have comparable rates to the $\pp$ transitions, which also
corresponds to a significant violation of HQSS. In the case of the
$\Uf\to\chi_{bJ}(1P)\omega$ transitions the HQSS prediction for the
relative yield of the $\chi_{bJ}(1P)$ states with different values of
$J$ also appears to be strongly
violated~\cite{Li:2014cia,Guo:2014qra}.

The fact that the states $\Ufo$, $\Uf$ and $\Us$ contain not only 
$b\bar{b}$ pairs, but also a ``molecular'' admixture of heavy mesons,
$B^{((*)*)}_{(s)}\bar{B}^{((*)*)}_{(s)}$, is crucial for understanding
their puzzling properties.
It was realized at the time of observation of the $\Ufo$, $\Uf$ and $\Us$
in 1980s~\cite{Lovelock:1985nb,Besson:1984bd} that too high splitting
between the $\Uf$ and $\Ufo$ states is due to the contribution of
hadron loops~\cite{Tornqvist:1984fx}, which is another language to
discuss the molecular admixture.
Enhanced transitions into hidden flavor final states are due to
rescattering of the on-shell heavy mesons (see
Fig.~\ref{fig:diagrams}~(right))~\cite{Meng:2007tk,Simonov:2008ci}.
Finally, the molecular admixture is also responsible for the violation
of HQSS. Indeed, an admixture of a specific $B \bar B$, $B \bar B^*$
or $B^* \bar B^*$ meson pair is not an eigenstate of the $b\bar{b}$
total spin. Table~\ref{tab:decomposition} presents the decomposition
of the $P$-wave $B^{(*)}\bar{B}^{(*)}$ pairs with $J^{PC}=1^{--}$ into
the $b\bar{b}$ spin eigenstates $\psi_{ij}$, where $i$ is the total
spin of the $b\bar{b}$ pair and $j$ is the total angular momentum
contributed by all other degrees of freedom, including both the spin
of light quarks and the orbital angular momentum
$L=1$~\cite{Kaiser:2002bm,Voloshin:2012dk}.
Various $\psi_{ij}$ components give rise to transitions that are
forbidden by HQSS for pure $b\bar{b}$ states. Experimental signatures
for $\psi_{ij}$ components are presented in
Table~\ref{tab:decomposition_decays}.
\begin{table}[tbh]
\caption{The decomposition of the $P$-wave $B^{(*)}\bar{B}^{(*)}$
  pairs with $J^{PC}=1^{--}$ into the $b\bar{b}$ spin eigenstates.}
\renewcommand{\arraystretch}{1.5} \centering
\begin{tabular}{ll}
\hline\hline
State & Decomposition into $b\bar{b}$ spin eigenstates \\
\hline
$B\bar{B}$ & $\frac{1}{2\sqrt{3}}\,\psi_{10} + \frac{1}{2}\,\psi_{11} +
\frac{\sqrt{5}}{2\sqrt{3}}\,\psi_{12} + \frac{1}{2}\,\psi_{01}$ \\
$B\bar{B}^*$ & $\frac{1}{\sqrt{3}}\,\psi_{10} + \frac{1}{2}\,\psi_{11} -
\frac{\sqrt{5}}{2\sqrt{3}}\,\psi_{12}$ \\
$(B^*\bar{B}^*)_{S=0}$ & $-\frac{1}{6}\,\psi_{10} - \frac{1}{2\sqrt{3}}\,\psi_{11} -
\frac{\sqrt{5}}{6}\,\psi_{12} + \frac{\sqrt{3}}{2}\,\psi_{01}$ \\
$(B^*\bar{B}^*)_{S=2}$ & $\frac{\sqrt{5}}{3}\,\psi_{10} -
\frac{\sqrt{5}}{2\sqrt{3}}\,\psi_{11} + \frac{1}{6}\,\psi_{12}$ \\
  \hline\hline
\end{tabular}
\label{tab:decomposition}
\end{table}
\begin{table}[tbh]
\caption{ Experimental signatures for the $b\bar{b}$ spin eigenstates
  $\psi_{ij}$. } \centering
\begin{tabular}{ll}
\hline\hline
Spin eigenstate & Expected decays \\
\hline
$\psi_{10}$ & $\Un\,\pp$, $\Un\,K^+K^-$ in $S$ wave \\
$\psi_{11}$ & $\Un\,\eta$, $\Un\,\eta'$ \\
$\psi_{11}$, $\psi_{12}$ & $\Un\,\pp$, $\Un\,K^+K^-$ in $D$ wave \\
$\psi_{01}$ & $\eta_b(nS)\,\omega$, $\eta_b(nS)\,\phi$, $\hbn\,\eta$, $\hbn\,\eta'$ \\
  \hline\hline
\end{tabular}
\label{tab:decomposition_decays}
\end{table}

One can expect that the closer is the physical state to the threshold,
the larger is the admixture of the corresponding meson pairs. Given
that the $\Ufo$ is only 20\,MeV above the $B\bar{B}$ threshold and
25\,MeV below the $B\bar{B}^*$ threshold, the $B\bar{B}$ should be the
dominant admixture in the $\Ufo$ wave
function~\cite{Voloshin:2012dk}. The $\psi_{10}$ component has the
same quantum numbers as a pure $b\bar{b}$ pair. It can contribute to
the $\Ufo\to\Uo\pp$ transitions, but since the phase space for the
$B\bar{B}$ pair is still small, the corresponding partial width is not
enhanced. The $\Ufo\to\Uo\eta$ transition could be due to the
$\psi_{11}$ component, thus the HQSS breaking finds a very natural
explanation. The $\psi_{11}$ and $\psi_{12}$ components should lead to
a $D$-wave contribution in the $\Ufo\to\Uo\pp$ decay, which was not
yet studied experimentally. Finally, the $\psi_{01}$ component could
contribute to the $\Ufo\to\hb\eta$ transitions. The same component
could result in the $\eta_b(1S)\omega$ final state, which has not yet
been studied experimentally.

In case of the $\Uf$ the dominant HQSS-violating contribution could be
the $B_s^*\bar{B}_s^*$ component, since the corresponding threshold is
only $40\,\mev$ below the $\Uf$ peak~\cite{Voloshin:2012dk}. In this
case one can expect that the $D$-wave component in the $\Uf\to\Uo
K^+K^-$ final state and the spin-singlet channels $\hbn\eta$,
$\hb\eta'$ and $\et\phi$ are enhanced. Only results on $\eta$
transitions were reported with relatively weak upper
limits~\cite{Tamponi:QWG2014}.
The distance to the $B^{(*)}\bar{B}^{(*)}$ thresholds of about
$260\,\mevm$ is relatively big compared to the splitting between the
thresholds of $45\,\mevm$, which possibly suppresses the contribution
of the non-strange $B$ meson pairs to the HQSS violation. Indeed,
there is no non-resonant contribution in the $\Uf\to\hbn\pp$
transitions.
The light quark content of the heavy meson admixture can be tested via
the ratio of the $\Uo\eta'$ vs. $\Uo\eta$ and $\hb\eta'$ vs. $\hb\eta$
decay rates~\cite{Voloshin:2012dk}.

The $\Us$ is situated near the $B_1(5721)\bar{B}$ threshold, where
$B_1(5721)$ is a narrow $P$-wave excitation with the spin-parity of the
light degrees of freedom $j^p=3/2^+$. A contribution of the
$B_1(5721)\bar{B}$ pairs to the $\Us$ decays has a very clear
experimental signature: the $\zbo\pi$ final state should be produced,
while the $\zbt\pi$ should not~\cite{Bondar:2016pox}. This prediction
is distinct from the observations at $\Uf$, where both $Z_b$ states
are produced in roughly equal proportion. Present data provide only a
very loose constraint on the relative yields of $\zbo$ and $\zbt$, and
do not exclude the single $\zbo$ hypothesis at high confidence
level~\cite{Abdesselam:2015zza}.

We conclude that hadronic transitions to hidden flavor channels
provide information on the parent bottomonium-like state.  Transitions
into bottomonia probe its heavy quark spin structure, while
transitions into molecular states directly probe a molecular admixture.

The experimental information on the decays of the $\Us$ is much less
detailed than that of the $\Uf$ due to a much smaller data sample of
effectively $3\,\fb$ compared to $121\,\fb$. Already these limited data
indicate that the decay pattern of the $\Us$ is different from that of the
$\Uf$ (see discussion in Ref.~\cite{Abdesselam:2015zza}), despite the
two states being relatively close in energy. The detailed comparison of
the two states is of interest and requires an increase of statistics
at the $\Us$, which can be realized at Belle-II. 

Energy dependence of the cross sections was measured only for the
$\Un\pp$ ($n=1,2,3$) and $\hbm\pp$ ($m=1,2$) final
states~\cite{Santel:2015qga,Abdesselam:2015zza}. The cross sections
exhibit no continuum contribution, and the parameters of the $\Uf$ and
$\Us$ peaks agree well among different channels (it is interesting to
confirm experimentally that the same is true for more hidden flavor
channels). Thus for a measurement of the relative rates of hadronic
transitions the on-resonance data are sufficient. The situation with
open flavor decays (like $B\bar{B}$, $B\bar{B}^*$,..)  is completely
different, since a significant continuum contribution is
seen~\cite{Aubert:2008ab,Santel:2015qga} and quite different
line-shapes are expected in different
channels~\cite{Ono:1985eu}. Therefore, studies of open flavor channels
require an energy scan; we discuss this in Section~\ref{sec:energy_scan}
below.

In the next two sections we consider the possibility of using the
hadronic transitions to search for missing conventional bottomonia and
for molecular states -- expected partners of the $\zbo$ and $\zbt$.

\subsection{Search for missing bottomonia below the $B\bar{B}$ threshold} 
\label{sec:missing_bottomonia}

The $121\,\fb$ data sample at the $\Uf$ was highly instrumental in
finding missing bottomonium levels. Belle observed the $\hb$ and
$\hbp$ states using $\Uf\to\hbn\pp$
transitions~\cite{Adachi:2011ji}. (We do not list these transitions in
Table~\ref{tab:partial_widths}, since they proceed entirely via the
intermediate $\zbo$ and $\zbt$ states~\cite{Belle:2011aa}.) Belle also
found first evidence for the $\ett$ state, precisely measured the
$\et$ mass and for the first time measured its width using prominent
radiative transitions from $\hbn$~\cite{Mizuk:2012pb}. In addition,
Belle observed the $\Uf\to\U_J(1D)\pp$ and $\Uf\to\U_J(1D)\eta$
transitions; the accuracy in the $\U_J(1D)$ mass is competitive with
measurements that use $\Uth\to\U_J(1D)\gamma$
transitions~\cite{PDG14}.

Bottomonium levels below the $B\bar{B}$ threshold that are still
missing are shown in Table~\ref{tab:transitions_to_missing_bb}.
\begin{table}[tbh]
\caption{ Missing bottomonium levels below the $B\bar{B}$ threshold,
  their quantum numbers, potential model predictions for
  masses~\cite{Godfrey:2015dia}, light hadrons emitted in the
  transitions from vector bottomonium-like states to the considered
  bottomonia and threshold of these transitions. }
\centering
\begin{tabular}{cccccl}
\hline\hline
Name & $L$ & $S$ & $J^{PC}$ & \hspace{8mm}Mass, $\mevm$\hspace{8mm} & Emitted hadrons [Threshold, $\gevm$] \\
\hline
$\eta_b(3S)$    & 0 & 0 & $0^{-+}$ & 10336 & $\omega$ [11.12], $\phi$ [11.36] \\
\hline
$h_b(3P)$       & 1 & 0 & $1^{+-}$ & 10541 & $\pp$ [10.82], $\eta$ [11.09], $\eta'$ [11.50] \\
\hline
$\eta_{b2}(1D)$ & 2 & 0 & $2^{-+}$ & 10148 & $\omega$ [10.93], $\phi$ [11.17] \\
\hline
$\eta_{b2}(2D)$ & 2 & 0 & $2^{-+}$ & 10450 & $\omega$ [11.23], $\phi$ [11.47] \\
$\U_J(2D)$     & 2 & 1 & $(1,2,3)^{--}$ & $10441-10455$ & $\pp$ [10.73], $\eta$ [11.00], $\eta'$ [11.41] \\
\hline
$h_{b3}(1F)$    & 3 & 0 & $3^{+-}$ & 10355 & $\pp$ [10.63], $\eta$ [10.90], $\eta'$ [11.31] \\
$\chi_{bJ}(1F)$ & 3 & 1 & $(2,3,4)^{++}$ & $10350-10358$ & $\omega$ [11.14], $\phi$ [11.38] \\
\hline
$\eta_{b4}(1G)$ & 4 & 0 & $4^{-+}$ & 10530 & $\omega$ [11.31], $\phi$ [11.55] \\
$\U_J(1G)$      & 4 & 1 & $(3,4,5)^{--}$ & $10529-10532$ & $\pp$ [10.81], $\eta$ [11.08], $\eta'$ [11.49] \\
\hline\hline
\end{tabular}
\label{tab:transitions_to_missing_bb}
\end{table}
The spin-singlet members of the $3S$, $3P$ and $1D$ multiplets are not
known, as well as the complete $2D$, $1F$ and $1G$ multiplets.
They can be searched for using transitions listed in
Table~\ref{tab:transitions_to_missing_bb}. Most of the thresholds for
final state particles are above $\Uf$ and $\Us$, thus one needs first
to find a new vector bottomonium-like state with sufficiently high
mass. However, quite a few final states are accessible already at the
$\Uf$ or $\Us$.

No result was reported yet for the $\Uf\to\U_J(2D)\pp$ transition. The 
chiral soft pion theorems ensure that the amplitude of a $\pp$ transition 
is bilinear in the momenta of the pions, independently of the structure of 
the vector bottomonium states [$\Uf$, $\Us$] such as presence of a molecular 
component, S - D mixing, etc. For this reason the transitions 
$\Uf\to\U_J(2D)\pp$ are strongly kinematically suppressed in comparison with 
$\Uf\to\U_J(1D)\pp$. Using the predictions of potential models for the mass 
of the $2^3D_2$ state about 10450\,MeV (see, e.g., Ref.~\cite{Kwong:1988ae}, 
one readily estimates that the rate of the $\Uf\to\U_J(2D)\pp$ carries a 
kinematical suppression factor of approximately $4 \times 10^{-3}$ as compared 
to  $\Uf\to\U_J(1D)\pp$. The relative kinematical suppression factor is not 
as dramatic at the higher energy of the $\Us$ peak: approximately 
$2.5 \times 10^{-2}$, but still is quite strong. For the case of 
$\eta$ transitions to the $\U_J(2D)$ bottomonium, only the $\Us$ peak is at 
or slightly above the kinematical threshold. Given that the $\eta$ transition 
is a $P$-wave process, a very strong kinematical suppression of the process 
$\Us \to \U_J(2D) \eta$ is to be expected. It is certainly possible that 
dynamical factors in the transition amplitudes do compensate, partially or 
fully, the kinematical suppression. However, realistically a search for 
the $\U_J(2D)$ bottomonium states would likely require exploring the 
$e^+ e^-$ annihilation at energies above the $\Us$ peak.

The production in $e^+e^-$ annihilation of the final states with the 
spin-singlet $h_{b3}(1F)$ and $\eta_{b4}(1G)$ bottomonium requires breaking 
of HQSS and is expected to be suppressed, unless such breaking is enhanced 
by molecular resonance effects, e.g. similar to the production of 
$h_b(mP) \pi \pi$ within the $\Uf$ and $\Us$ resonances due to intermediate 
exotic $Z_b$ molecular states. Furthermore, the light mesons produced in 
association with either spin-singlet or spin-triplet $1F$ or $1G$ bottomonium 
should carry a large angular momentum, resulting in a kinematical suppression 
of the yield near the thresholds. A possible exception from this threshold 
behavior may be applicable to the production of the final state 
$\chi_{b2}(1F) \omega$ [or  $\chi_{b2}(1F) \phi$], where an $S$-wave process 
is kinematically allowed. However, a presence of an $S$-wave requires 
breaking of HQSS, so that an observation of such process would definitely 
provide an insight into the mechanisms for violation of HQSS.

The transitions listed in Table~\ref{tab:transitions_to_missing_bb}
can be reconstructed inclusively using missing mass of the emitted
light hadrons. In case of the spin-triplet levels there are also final
states convenient for exclusive reconstruction. The dominant
transitions between the bottomonia below the $B\bar{B}$ threshold are
radiative $E1$ transitions that change orbital angular momentum of the
$b\bar{b}$ pair by one unit and conserve the $b\bar{b}$ spin. Thus the
chain
$\U_J(1G)\to\gamma\chi_{bJ}(1F)\to\gamma\gamma\U_J(1D)\to\gamma\gamma\gamma\chi_{bJ}(1P)\to\gamma\gamma\gamma\gamma\Uo$
corresponds to dominant transitions and can be used for exclusive
reconstruction with $\Uo\to\ee$ or $\uu$. More details on the
bottomonium decays can be found in e.g. Ref.~\cite{Godfrey:2015dia}.

\subsection{Search for molecular states -- partners of the $\zbo$ and
  $\zbt$ }
\label{sec:zb_partners}

In the single pion transitions from the $\Uf$ and $\Us$ Belle observed
isovector bottomonium-like states $\zbo$ and
$\zbt$~\cite{Belle:2011aa}. The Breit-Wigner masses of the $\zbo$ and
$\zbt$ are located within the experimental uncertainty of about
$2\,\mevm$ at the $B\bar{B}^*$ and $B^*\bar{B}^*$ thresholds,
respectively. These states are known to decay to $\Un\pi$ ($n=1,2,3$)
and $\hbm\pi$ ($m=1,2$) channels; however, the dominant decays with
about 80\% branching fractions are $\zbo\to B\bar{B}^*$ and
$\zbt\to{B}^*\bar{B}^*$~\cite{Garmash:2015rfd}. Such a decay pattern
is a ``smoking gun'' of the molecular structure, $B\bar{B}^*$ for the
$\zbo$ and $B^*\bar{B}^*$ for the
$\zbt$~\cite{Bondar:2011ev}. Measured spin and parity of $J^P=1^+$ for
both states~\cite{Garmash:2014dhx} correspond to $B\bar{B}^*$ and
$B^*\bar{B}^*$ in the $S$-wave, which supports the molecular
interpretation. Recent combined analysis of the $B^{(*)}\bar{B}^*$ and
$\hbm\pi$ channels using phenomenologically motivated expressions for
amplitudes indicates that $\zbo$ and $\zbt$ may in fact be virtual
molecular states with poles within $2\,\mev$ from the corresponding
thresholds~\cite{Hanhart:2015cua,Guo:2016bjq}.

In the limit of exact HQSS the mechanism that binds $B$ mesons to form
the $Z_b$ states is determined by light degrees of freedom, while the
total spin of the $b\bar{b}$ pair plays a classification role only:
rotating this spin one can find other molecular states that are
partners of the
$Z_b$'s~\cite{Voloshin:2011qa}. Table~\ref{tab:zb_partners} gives the
predicted states with their composition and quantum numbers.
\begin{table}[tbh]
\caption{ Expected molecular states with the structure $B\bar{B}$,
  $B\bar{B}^*$ and $B^*\bar{B}^*$. } \centering
\begin{tabular}{ccccl}
\hline\hline
$I^G(J^P)$ & Name & Composition & Co-produced particles [Threshold, $\gevm$] & Decay channels \\
\hline
$1^+(1^+)$ & $Z_b$     & $B\bar{B}^*$   & $\pi$ [10.75] & $\Un\pi$, $\hbn\pi$, $\eta_b(nS)\rho$ \\
$1^+(1^+)$ & $Z_b'$    & $B^*\bar{B}^*$ & $\pi$ [10.79] & $\Un\pi$, $\hbn\pi$, $\eta_b(nS)\rho$ \\
$1^-(0^+)$ & $W_{b0}$  & $B\bar{B}$     & $\rho$ [11.34], $\gamma$ [10.56] & $\Un\rho$, $\eta_b(nS)\pi$ \\
$1^-(0^+)$ & $W_{b0}'$ & $B^*\bar{B}^*$ & $\rho$ [11.43], $\gamma$ [10.65] & $\Un\rho$, $\eta_b(nS)\pi$ \\
$1^-(1^+)$ & $W_{b1}$  & $B\bar{B}^*$   & $\rho$ [11.38], $\gamma$ [10.61] & $\Un\rho$ \\
$1^-(2^+)$ & $W_{b2}$  & $B^*\bar{B}^*$ & $\rho$ [11.43], $\gamma$ [10.65] & $\Un\rho$ \\
\hline
$0^-(1^+)$ & $X_{b1}$  & $B\bar{B}^*$   & $\eta$ [11.15] & $\Un\eta$, $\eta_b(nS)\omega$ \\
$0^-(1^+)$ & $X_{b1}'$ & $B^*\bar{B}^*$ & $\eta$ [11.20] & $\Un\eta$, $\eta_b(nS)\omega$ \\
$0^+(0^+)$ & $X_{b0}$  & $B\bar{B}$     & $\omega$ [11.34], $\gamma$ [10.56] & $\Un\omega$, $\eta_b(nS)\eta$ \\
$0^+(0^+)$ & $X_{b0}'$ & $B^*\bar{B}^*$ & $\omega$ [11.43], $\gamma$ [10.65] & $\Un\omega$, $\eta_b(nS)\eta$ \\
$0^+(1^+)$ & $X_b$     & $B\bar{B}^*$  & $\omega$ [11.39], $\gamma$ [10.61] & $\Un\omega$ \\
$0^+(2^+)$ & $X_{b2}$  & $B^*\bar{B}^*$ & $\omega$ [11.43], $\gamma$ [10.65] & $\Un\omega$ \\
  \hline\hline
\end{tabular}
\label{tab:zb_partners}
\end{table}
All the expected, but not yet observed, isovector states have negative
$G$-parity, which is opposite to that of the $Z_b$ resonances. For
this reason they can not be produced in $e^+e^-$ annihilation with the
emission of a single pion. The most natural channel for their
production is the emission of $\rho(770)$ meson.  For the transitions
from $\Uf$ the maximal $\pp$ invariant mass is only $300\,\mevm$ which
makes such a production, due to the low-mass `tail' of the $\rho$
resonance, strongly suppressed. At the $\Us$ the maximal mass is
$440\,\mevm$, thus there are better chances to observe the
$W_{b0}$. An alternative way to produce the electrically neutral
isotopic components of these molecular states at the $\Uf$ is provided
by emission of a photon.  Naturally, the rates of such radiative
processes carry the suppression factor of $\alpha$.

No isosinglet states were seen yet in the bottomonium sector, and at present 
they are purely hypothetical. It has to be mentioned that only the isovector 
states in the Table~\ref{tab:zb_partners} are related to the observed $Z_b$ 
resonances. For the isosinglet ones the interaction of the light components 
is different, and it is not known whether this interaction results in 
near-threshold singularities for the heavy meson-antimeson pairs. 
Furthermore, the isosinglet states are affected by mixing with a pure 
$b \bar b$ bottomonium, which mixing can result in yet not fully predictable 
modification of their properties. In particular, the admixture of a compact 
$b \bar b$ component should result in that the isoscalar states, unlike 
those with $I=1$, can be produced in hard processes, e.g. at LHC. 
An example of such a behavior in the charmonium sector is the $X(3872)$ 
resonance. Being essentially a threshold singularity in the 
$D^0 \bar D^{*0} + \bar D^0 D^{*0}$ channel, it apparently contains a 
short-distance $c \bar c$ charmonium core through which it is produced in 
hard processes: the $B$ meson decays and at high-energy 
proton-(anti)proton colliders (see, e.g., in the review~\cite{Voloshin:2007dx}).

There is also a good reason to avoid deducing the existence and the 
properties of the hypothetical $J^{PC}=1^{++}$ isoscalar bottomonium-like 
resonance $X_b$ from those of the $X(3872)$. Namely, in the latter the 
isotopic symmetry is badly broken by the isotopic mass differences of the 
$D$ mesons, so that the $X(3872)$ is a strong mixture of $I=0$ and 
$I=1$ states. For the $B$ mesons the isotopic mass differences are very small, 
and the separation between isoscalar and isovector states should be very 
well preserved. 

The isosinglet states can be produced in $\eta$ or $\omega$
transitions (depending on the $C$-parity), however, in this case the
$\Us$ energy is insufficient, and one has to find a higher vector
state with the mass above $11.43\,\gevm$.

Details of the interaction resulting in the existence of the $Z_b$ peaks 
are not yet understood. One might speculate at this point that if the 
dominant one is the interaction between the gluonic degrees of freedom 
in the $B^{(*)}$ mesons, some form of flavor SU(3) symmetry may be present 
between the molecular states giving rise to existence of strange analogs 
of the $Z_b$ resonances. Such bottomonium-like strange resonances, 
$Z_{bs}$ related to the combinations of the channels $B_s^* \bar B$ and 
$B_s \bar B^*$ and also to $B_s^* \bar B^*$, should have masses, respectively, 
near 10695\,MeV and 10740\,MeV, and can be produced in $e^+e^-$ annihilation 
in association with a kaon: $e^+e^- \to Z_{bs} K$, at energies above 
11.20\,GeV. These resonances would decay into the states of bottomonium plus 
a kaon, and also to heavy meson pairs with one $B$ meson being either 
$B_s$ or $B_s^*$. By any measure, an observation of such bottomonium-like 
molecules with strangeness would be of paramount importance for studies 
of hadronic dynamics.

\section{Energy scan of the cross sections}
\label{sec:energy_scan}

Measurement of the total hadronic cross section does not require high
statistics. In recent energy scans BaBar and Belle used $25\,\pb$ and
$50\,\pb$ per point,
respectively~\cite{Aubert:2008ab,Santel:2015qga}. Much more data are
necessary to measure exclusive cross sections. Belle performed a
high-statistics energy scan with roughly $1\,\fb$ per point and
measured the $\Un\pp$ ($n=1,2,3$), $\hbm\pp$ ($m=1,2$) and
$B_s^{(*)}\bar{B}_s^{(*)}$ cross
sections~\cite{Santel:2015qga,Abdesselam:2015zza,Abdesselam:2016tbc}. Results
for $B\bar{B}$, $B\bar{B}^*$, $B^*\bar{B}^*$ and
$B^{(*)}\bar{B}^{(*)}\pi$ are still expected~\cite{Mizuk:QWG2016}. The
statistical uncertainty of the measurements is rather large, thus at
Belle-II it is useful to collect about $10\,\fb$ per scan point.
Given that expected energy smearing at Belle-II is similar to that at
Belle and thus is close to 5\,MeV, no narrow peak will be missed if
the step of the scan is 10\,MeV.
The total hadronic cross section is measured up to 11.2\,GeV, while
the exclusive cross sections are measured up to 11.02\,GeV. The
maximal energy of the SuperKEKB collider, limited by the injection
system, is 11.24\,GeV~\cite{Ohnishi:2013fma}.

\subsection{Cross sections for open flavor channels}

\begin{figure}[htbp]
\centering
\includegraphics[width=0.52\linewidth]{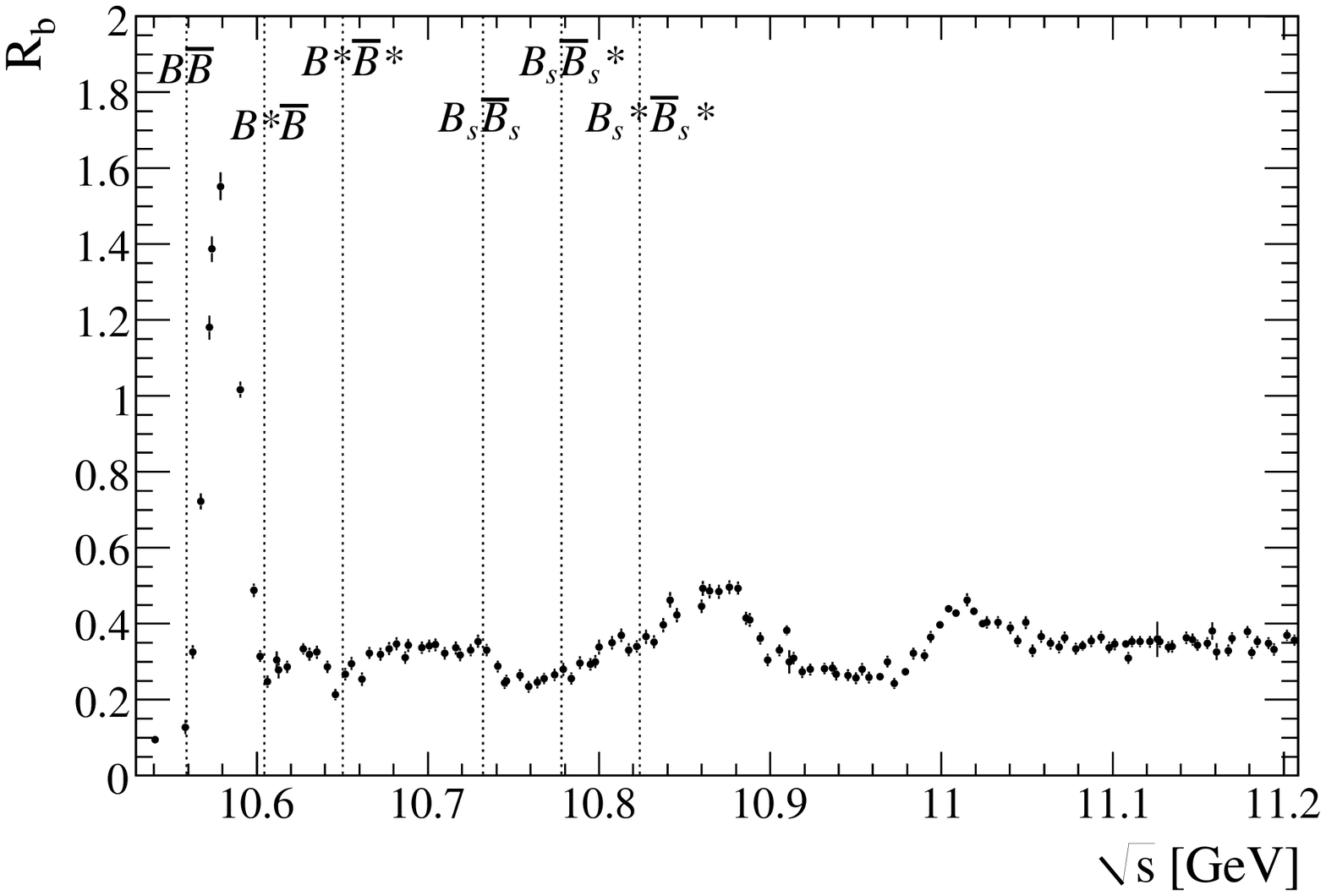}\hfill
\includegraphics[width=0.45\linewidth]{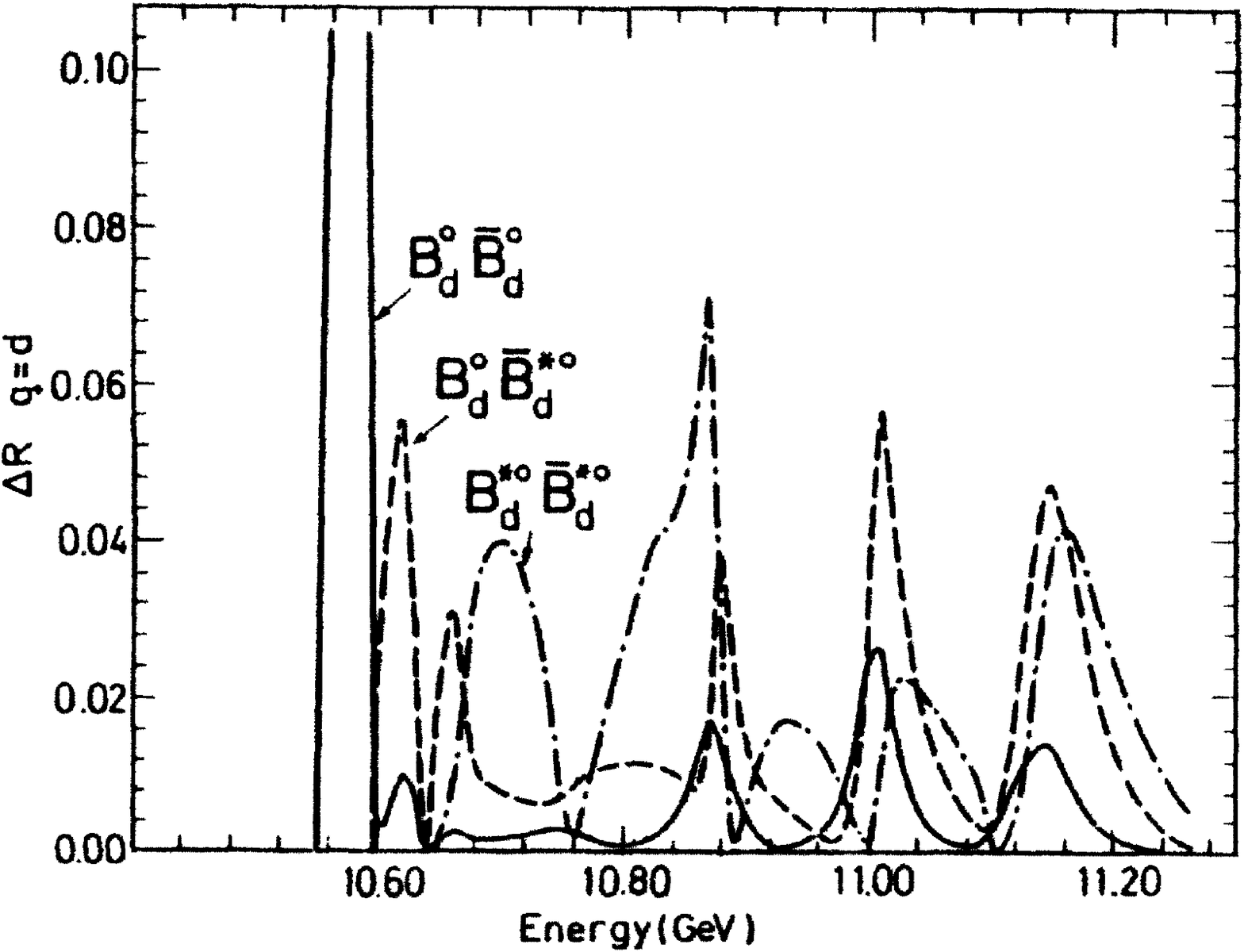}
\caption{ The $R_b$ scan results from BaBar~\cite{Aubert:2008ab}
  (left), and the expected in Unitarized Quark Model contributions of
  the $B\bar{B}$, $B\bar{B}^*$ and $B^*\bar{B}^*$
  channels~\cite{Ono:1985eu} (right). }
\label{fig:UQM}
\end{figure}

The measured total hadronic cross section, usually presented as
$R_b=\sigma(\ee\to b\bar b) / \sigma^0_{\mu\mu}$, where
$\sigma^0_{\mu\mu}$ is the Born $\ee\to\uu$ cross section, has several
features (see Fig.~\ref{fig:UQM}~(left)): these are peaks of
the $\Ufo$, $\Uf$ and $\Us$, and dips near the $B\bar{B}^*$,
$B^*\bar{B}^*$ and $B_s^*\bar{B_s}^*$
thresholds~\cite{Santel:2015qga,Aubert:2008ab}.
The channels with the bottomonium in the final state contribute only a
few percent to the total $b\bar{b}$ cross section; the rest is due to
the open flavor channels, such as $B\bar{B}$, $B\bar{B}^*$,
$B^*\bar{B}^*$, $B\bar{B}\pi$, $B\bar{B}^*\pi$, $B^*\bar{B}^*\pi$,
$B_s\bar{B}_s$, $B_s\bar{B}_s^*$, $B_s^*\bar{B}_s^*$ etc. The
corresponding exclusive cross sections are expected to have
significantly more features than their sum. As an example,
Fig.~\ref{fig:UQM}~(right) shows the predictions of the Unitarized
Quark Model~\cite{Ono:1985eu}.
The oscillatory behavior of exclusive cross sections in this model is
due to the nodes of the $\Uf$ and $\Us$ wave functions.
The individual cross sections contain considerably more information
than their sum. Their measurements are extremely important for
understanding this energy region. In particular, they will allow to
determine the $\ee$ widths of $\Uf$ and $\Us$, that are needed to
determine their branching fractions. Moreover, a combined analysis of
the energy dependence of the open-flavor and hidden-flavor cross
sections in the framework of the coupled-channel models is crucial for
determining their most essential basic properties, such as the
positions of the poles and inter-channel couplings. Thus, the studies
along these lines could establish the picture of the vector
bottomonium-like states as superpositions of a pure $b\bar{b}$
bottomonium and a molecular $B_{(s)}^{(*)}\bar{B}_{(s)}^{(*)}$
component.

\subsection{Cross sections for hidden-flavor channels}

Hidden-flavor cross sections provide input for the coupled-channel
analysis and play important role in searching for new states.

One alternative to the molecular picture, currently discussed in the
literature, is the diquark-antidiquark model~\cite{Ali:2009es}, where
the quarks and antiquarks making up a four-quark state are the same as
in a molecule, however they are arranged (or clustered) differently:
the two quarks form a diquark, while the two antiquarks form an
antidiquark. Both diquark and antidiquark are colored objects,
therefore such a system is very compact contrary to meson-antimeson
molecule. The (anti-)quarks within the (anti-)diquark are more tightly
bound than the diquark and anti-diquark. This model can also explain
the OZI rule violation and the HQSS breaking, however, the decays to
open-flavor channels should not dominate, since the diquark is broken
both in open-flavor decays and in transitions to bottomonia, while for
the open-flavor decays the phase space is much smaller. Thus the $\Uf$
and $\Us$ states are not good candidates for compact tetraquarks,
since their signals are seen in the total hadronic cross section, and
thus the open-flavor channels are likely to give a dominant
contribution to the $\Uf$ and $\Us$ peaks. A unique signature of the
tetraquark is a peak in the cross section of $\ee$ annihilation into
bottomonia and light hadrons without a significant peak in the total
hadronic cross section. At the moment there is no experimental
evidence for such candidates; the hints for the $Y_b(10890)$ suggested
in Ref.~\cite{Ali:2009es} and $Y_b(10900)$~\cite{Ali:2009pi} were not
confirmed with larger data samples~\cite{Santel:2015qga}. A search
for the compact tetraquarks remains intriguing and important topic.

In the charmonium sector there are states that exhibit themselves as
peaks in hidden-flavor channels (e.g., $\jp\pp$, $\psp\pp$, $h_c\pp$),
but are not seen in the total hadronic cross section, or even in
exclusive open-charm final states~\cite{PDG14}. The nature of these
charmonium-like states is not yet understood; given that each peak is
dominantly seen in only one final state, the notion of hadrocharmonium
has been proposed~\cite{Dubynskiy:2008mq}. This ``selectivity'' of
final states makes hadrocharmonium distinct from a compact
tetraquark. There could exist a $b\bar b$ partner of the
hadrocharmonium, the hadrobottomonium.

The states with molecular admixture should be seen in the open-flavor
channels or in the total hadronic cross section, as is the case for the
$\Ufo$, $\Uf$ and $\Us$, since dominance of the open-flavor channels
is a ``smoking gun'' of the molecular structure. Nevertheless, 
hidden-flavor channels could have better sensitivity because of higher
reconstruction efficiency. Moreover, in open-flavor channels there is
always a continuum contribution, which is zero in hidden-flavor
channels in all known so far cases.

Thus, the hidden-flavor channels could have better sensitivity than
the open-flavor ones in searches for states with molecular admixture;
they are unique for exotic states with different than molecular
structures, such as compact tetraquarks or hadrobottomonia. The final
states to be investigated are the same as already found at the $\Uf$ or
proposed to be searched for (see Tables~\ref{tab:partial_widths} and
\ref{tab:decomposition_decays}).

\subsection{Promising $\ee$ energy regions}

Molecular states are naturally located (and produce the largest
effects) near the corresponding threshold. The positions of the
thresholds are listed in Table~\ref{tab:thresholds}, where we consider
only narrow $S$- and $P$-wave mesons and baryons.
\begin{table}[tbh]
\caption{ Thresholds of narrow $S$ and $P$ wave mesons and baryons. }
\centering
\begin{tabular}{cc}
\hline\hline
Particles & Threshold, $\gevm$ \\
\hline
$B^{(*)}\bar{B}^{**}$ & 11.00 -- 11.07 \\
$B_s^{(*)}\bar{B}_s^{**}$ & 11.13 -- 11.26 \\
$\Lambda_b\,\bar{\Lambda}_b$ & 11.24 \\
$B^{**}\bar{B}^{**}$ & 11.44 -- 11.49 \\
$B_s^{**}\bar{B}_s^{**}$ & 11.48 -- 11.68 \\
$\Lambda_b\,\bar{\Lambda}_b^{**}$ & 11.53 -- 11.54 \\
$\Sigma_b^{(*)}\,\bar{\Sigma}_b^{(*)}$ & 11.62 -- 11.67 \\
$\Lambda_b^{**}\,\bar{\Lambda}_b^{**}$ & 11.82 -- 11.84 \\
\hline\hline
\end{tabular}
\label{tab:thresholds}
\end{table}
The regions of $B^{(*)}\bar{B}^{**}$ and $B_s^{(*)}\bar{B}_s^{**}$ are
basically within the current reach of the SuperKEKB collider of
11.24\,GeV. An increase of the energy by at least 100\,MeV will give
a possibility to explore the $\Lambda_b\,\bar{\Lambda}_b$ threshold
region and to search for baryon-antibaryon molecular states. Such
states are almost certainly there, as can be judged from the
charmonium sector~\cite{Pakhlova:2008vn}.

It is also quite important to find a high-mass state to serve as a
source of missing bottomonia below the $B\bar{B}$ threshold and of
molecular states -- partners of the $\zbo$ and $\zbt$, as discussed in
Sections~\ref{sec:zb_partners} and \ref{sec:missing_bottomonia}. The
thresholds for various transitions are listed in
Tables~\ref{tab:transitions_to_missing_bb} and
\ref{tab:zb_partners}. The maximal threshold is at $11.55\,\gev$. In
order to cover the molecular states at the $B_s^{(*)}\bar{B}_s^{(*)}$
thresholds the peak should be above $11.61\,\gev$. This is the region
of $B^{**}\bar{B}^{**}$, $B_s^{**}\bar{B}_s^{**}$,
$\Lambda_b\,\bar{\Lambda}_b^{**}$ and
$\Sigma_b^{(*)}\,\bar{\Sigma}_b^{(*)}$ thresholds. Finally,
investigation of the complete region of resonances would require to
increase the energy up to 12\,GeV.

In the next section we discuss the $B_s^{(*)}\bar{B}_s^{**}$
threshold, which is within the current reach of SuperKEKB.

\subsubsection{Search for narrow excited $0^+$ and $1^+$ strange $B_s$ mesons}

Based on the heavy quark symmetry one should expect existence in the 
$b$-flavored sector of analogs of the charmed strange $D_{s0}(2317)$ and 
$D_{s1}(2460)$ mesons: $B_{s0}$ and $B_{s1}$. Their expected masses can be 
estimated within the heavy quark expansion as 5715\,MeV and 5763\,MeV, 
respectively. These mesons can be produced in $e^+e^-$ annihilation in 
pairs with the ground-state strange $B_s$ and $B^*_s$ mesons. Moreover, 
an $S$-wave production is allowed for the pairs 
$B_{s0} \bar B^*_s + c.c.$, $B_{s1} \bar B_s + c.c.$ and $B_{s1} \bar B^*_s + c.c.$. The thresholds for the former two channels practically coincide at 
the c.m. energy of 11.13\,GeV, while for the latter the threshold is expected 
to be at approximately 11.18\,GeV, and one can expect, due to a presence 
of the $S$-wave, a measurable production cross section at energies just 
above the threshold. The final states in these processes necessarily 
contain $B_s$ and $\bar B_s$ mesons, and the excited $B_{s0}$ and $B_{s1}$ 
strange bottom mesons can be sought for by studying the spectrum of the 
invariant mass recoiling against a reconstructed $B_s$ meson. 

Another interesting possibility may potentially arise if there is a 
molecular $S$-wave resonance, $\Upsilon_{s \bar s}$, near the threshold of 
the $B_{s0} \bar B^*_s + c.c.$ and $B_{s1} \bar B_s + c.c.$ pairs, i.e. 
near 11.13\,GeV. An existence of such a resonance (or a double resonance) 
can be affected by the coincidence of the thresholds for the two types of 
pairs (which is not the case for the final state $B_{s1} \bar B^*_s + c.c.$). 
One can expect that such a resonance, in addition to decay channels with 
open $b$ flavor, would have a small, but possibly measurable, 
branching fraction for decay into bottomonium and light mesons:  e.g. 
$\Upsilon_{s \bar s} \to \Upsilon(nS) K \bar K$ (with $n=1,\,2$), 
or  $\Upsilon_{s \bar s} \to \chi_{bJ} (1P) \phi$. [The former decay is 
somewhat similar to the part of decays $\Uf \to \Upsilon(nS) \pi \pi$ not associated with the $Z_b$ resonances. The branching fraction for this part is in 
the ballpark of (a few)$\times 10^{-3}$.] Thus a scan of these channels 
with hidden strangeness near the c.m. energy of 11.13\,GeV can 
reveal existence of molecular resonances with the excited strange bottom 
mesons\footnote{By applying the same picture in the charm sector, one can 
expect a similar phenomenon due to the $S$-wave production in $e^+e^-$ 
annihilation of the pairs $D_{s0}(2317) \bar D^*_s + c.c.$ and 
$D_{s1}(2460) \bar D_s + c.c.$ at and/or above their common threshold 
at 4.43\,GeV. The available data of a study using initial state radiation at Belle~\cite{Shen:2014gdm} indicate a certain activity in the final state $J/\psi K \bar K$ above 4.4\,GeV with possible structures. However  more detailed data  on 
the yield of strange $D_s$ mesons and/or of  
$J/\psi K \bar K$ in that energy range are needed for further conclusions.}. It can be also mentioned that 
$S$-wave decays into $h_b(mP) \eta$ are also allowed for 
$\Upsilon_{s \bar s}$, in analogy with the decay 
$\Upsilon(4S) \to h_b(1P) \eta$, although it is not clear at present, 
how strong the breaking of HQSS, that is required for such decays, is. 
However, if these transitions from $\Upsilon_{s \bar s}$ proceed at a 
measurable rate, they can include the decay 
$\Upsilon_{s \bar s} \to h_b(3P) \eta$ and thus provide a unique gateway 
to studies of the $h_b(3P)$ state of bottomonium.

\section{Conclusions}

All hidden-beauty hadrons above the $B\bar{B}$ threshold have
properties inconsistent with their interpretation as pure $b\bar{b}$
bottomonia. The vector bottomonium-like states $\U(10580)$,
$\U(10860)$ and $\U(11020)$ [or the $\Ufo$, $\Uf$ and $\Us$] are likely
mixtures of conventional $b\bar{b}$ bottomonia and pairs of $B\bar{B}$
or $B_s\bar{B}_s$ mesons in ground or excited states. The isospin-one
axial states $\zbo$ and $\zbt$ likely have purely molecular structures
of $B\bar{B}^*$ and $B^*\bar{B}^*$, respectively. We propose a
dedicated data taking program that will establish the above
interpretation, check its predictions and search for new bottomonium
and bottomonium-like states.

We propose to perform a high-statistics energy scan from the
$B\bar{B}$ threshold up to the highest possible energy, with $10\,\fb$
per point and $10\,\mev$ step. These data will allow to measure cross
sections of $\ee$ annihilation into exclusive open-flavor and
hidden-flavor final states, such as $B\bar{B}$, $B\bar{B}^*$,
$B^*\bar{B}^*$, $B\bar{B}\pi$, $B\bar{B}^*\pi$, $B^*\bar{B}^*\pi$,
$B_s\bar{B}_s$, $B_s\bar{B}_s^*$, $\Un\pp$, $\Un\eta$, $Z_b\pi$ and
others. Combined analysis of these cross sections using
coupled-channel framework will determine all basic properties of the
vector bottomonium-like states, such as their pole positions and
interchannel couplings. These results are crucial for establishing the
nature of the vector bottomonium-like states. The hidden-flavor
channels will be used to search for states that do not exhibit
themselves in the total cross section or in open-flavor channels;
among such states could be compact tetraquarks and hadroquarkonia.

The states with molecular admixture are naturally located near the
corresponding threshold. The present energy limit of the SuperKEKB
accelerator of 11.24\,GeV will allow to investigate the
$B^{(*)}\bar{B}^{**}$ and $B_s^{(*)}\bar{B}_s^{**}$ threshold
regions. Increase of maximal energy by at least 100\,MeV will allow to
explore the $\Lambda_b\,\bar{\Lambda}_b$ threshold region and search 
for baryon-antibaryon molecular states, which should be there,
as can be judged from the charmonium sector. The region of promising
thresholds extends up to 12\,GeV.

Observation of new vector states is of utmost importance for studies
of molecular states -- partners of the $\zbo$ and $\zbt$. Transitions from
high mass bottomonium-like states provide a unique way to produce purely
molecular states. Energy above 11.5\,GeV is of special interest, as
most of the relevant hadronic transitions become kinematically allowed. 
Thus searches for the molecular states provide additional motivation for
increase of the SuperKEKB energy up to at least 11.5 -- 11.6\,GeV.

At each new vector state we propose to collect $500\,\fb$ to perform a
detailed study of corresponding transitions, since they provide
information on the structure of the state, to search for missing
conventional bottomonia and to search for molecular states. A region
near the $B_s^{(*)}\bar{B}_s^{**}$ threshold is also of interest to
search for missing $P$-wave $B_s$ mesons.

Available data at $\Us$ are limited to $5\,\fb$ that were taken at five
scan points, with the effective luminosity of $3\,\fb$. Only the most
prominent transitions $\Us\to\Un\pp$ and $\Us\to Z_b\pi$ are
known. An observed pattern of transitions is somewhat different from that
at the $\Uf$, despite that the two states are close in energy. More detailed
comparison requires an increase of statistics at the $\Us$ and could
provide interesting information on the structure of both states. Given
that many transitions were observed at the $\Uf$ and there are still
channels to be investigated, the list of measurements to be performed
at the $\Us$ is quite long. Even non-observation of some channels will be
of interest for comparison of the $\Uf$ and $\Us$. The $\Us$ data will
provide a rich physics output even for relatively small ammount of a
few tens of $\fb$. This makes $\Us$ an attractive option for the
Belle-II initial data taking.

\section*{Acknowledgments}

We are grateful to Simon Eidelman for valuable comments. The work of
M.B.V. is supported in part by U.S. Department of Energy Grant
No.\ DE-SC0011842 and was performed in part at the Aspen Center for
Physics, which is supported by National Science Foundation grant
PHY-1066293. R.V.M. acknowledges support from the Russian Science
Foundation (Grant No. 15-12-30014).

\end{document}